\documentclass[a4paper]{article}

\usepackage{INTERSPEECH2022}
\usepackage{tablefootnote}
\usepackage{threeparttable}
\usepackage{multirow,tabularx}
\usepackage{adjustbox}
\usepackage{cite}
\usepackage[stable]{footmisc}
\title{Towards Improved Zero-shot Voice Conversion with Conditional DSVAE}
\name{Jiachen Lian$^{1,2, *}$\thanks{$*$ Equal Contribution. Work done when Jiachen was an intern at Tencent AI Lab, Bellevue, WA}, Chunlei Zhang$^{2,*}$,  Gopala Krishna Anumanchipalli$^{1}$, Dong Yu$^{2}$}
  
\address{$^{1}$ UC Berkeley, EECS, CA $^{2}$ Tencent AI Lab, Bellevue, WA}
\email{jiachenlian@berkeley.edu, cleizhang@tencent.com, gopala@berkeley.edu, dyu@tencent.com}

\begin{document}

\maketitle
\begin{abstract}
Disentangling content and speaking style information is essential for zero-shot non-parallel voice conversion (VC). Our previous study investigated a novel framework with disentangled sequential variational autoencoder (DSVAE) as the backbone for information decomposition. We have demonstrated that simultaneous disentangling
content embedding and speaker embedding from one utterance is feasible for zero-shot VC. In this study, we continue the direction by raising one concern about the prior distribution of content branch in the DSVAE baseline. We find the random initialized prior distribution will force the content embedding to reduce the phonetic-structure information during the learning process, which is not a desired property. Here, we seek to achieve a better content embedding with more phonetic information preserved. We propose \textbf{conditional DSVAE}, a new model that enables content bias as a condition to the prior modeling and reshapes the content embedding sampled from the posterior distribution. In our experiment on the VCTK dataset, we demonstrate that content embeddings derived from the conditional DSVAE overcome the randomness and achieve a much better phoneme classification accuracy, a stabilized vocalization and a better zero-shot VC performance compared with the competitive DSVAE baseline. 

\end{abstract}
\section{Introduction}
Voice Conversion (VC) is a technique that converts the non-linguistic information of a given utterance to a target style (e.g., speaker identity, emotion, accent or rhythm etc.), while preserving the linguistic content information. VC has become a very active research topic in speech processing with potential applications in privacy protection speaker de-identification, audio editing or sing voice conversion/generation \cite{sisman2020overview,bahmaninezhad2018convolutional,zhang2020durian, deep-fake}.

%Based on the conditions of source and target speakers that a VC system can access in the training phase, we can categorise current VC approaches into one-to-one, many-to-one, many-to-many, any-to-many and any-to-any (i.e., zero-shot) VC. Conventional studies focus on one-to-one VC, which requires parallel data between a pair of source-target speakers. In these systems, acoustic features of source and target utterances are firstly aligned frame-wise with an alignment module, and a conversion transformation is trained to map time-aligned source acoustic features to target features. The requirement of parallel data limits the application of such models in the real world. Recent

Current VC systems embrace the technological advancements from statistical modeling to deep learning and have made a major shift on how the pipeline develops~\cite{sisman2020overview}. For example, the conventional VC approaches with parallel training data utilize a conversion module to map source acoustic features to target acoustic features, the source-target pair has to be aligned before the mapping~\cite{berndt1994using}. With the advent of sequence-to-sequence models even without the alignment prerequisite, better VC performance is reported~\cite{zhang2019sequence}. For VC with non-parallel data, direct feature mapping method is difficult. Instead, studies start to explicitly learn the speaking style and content representations and train a neural network as a decoder to reconstruct the acoustic feature, with the assumption that the decoder can also generalize well when the content and speaker style is swapped during the conversion. Among the approaches, phonetic posteriorgrams (PPGs) and pre-trained speaker embeddings are widely used as the content and speaking style representations \cite{sun2016phonetic,liu2018voice,guo2020phonetic,zhang2021transfer}. However, developing such system usually requires a big amount of external data with rich transcriptions and speaker labels. The relatively small-footprint AUTOVC and AdaIN-VC employ encoder-decoder frameworks for zero-shot VC~\cite{autovc,adaIN}. The encoder decomposes the speaking style and the content information into the latent embedding, and the decoder generates a voice sample by combining both disentangled information. Nevertheless, these models require supervisions such as positive pair of utterances (i.e., two utterances come from the same speaker), and the systems still have to rely on pre-trained speaker models. Progress has also been made with generative adversarial networks (GAN) based VC systems~\cite{kameoka2018stargan,kaneko2018cyclegan,Li2021StarGANv2VCAD, li2020cvc}. This categorical of method usually assumes that the speaker of source-target VC pair is pre-known, which limits the application of such models in the real world. At the same time, bunch of regularization terms have to be applied in the training process, which imposes generalization doubts to such systems for zero-shot non-parallel VC scenarios.  

Our previous study proposed a novel disentangled sequential variational autoencoder (DSVAE)~\cite{D-DSVAE} as a backbone framework for zero-shot non-parallel VC. We designed two branches in the encoder of DSVAE to hold the time-varying and the time-invariant components, where balanced content and speaking style information flow is achieved with the VAE training \cite{vae}. We demonstrated that the vanilla VAE \cite{vae, dsvae} loss can be extended to force strong disentanglement between speaker and content components, which is essential for the success of challenging zero-shot non-parallel VC.

In this study, we continue the direction by further improving the disentangled representation learning in the DSVAE framework. One major concern is raised after we analyzed the content embedding learned from our DSVAE baseline~\cite{D-DSVAE}. We find that the random initialed prior distribution in the content branch of the baseline DSVAE is not optimal to preserve the phonetic/content structure information. The randomness of content embedding $z_c$ has a negative impact to phoneme classification and VC. To cope with this issue, we propose $conditional$ DSVAE (C-DSVAE), an improved framework that corrects the randomness in the content prior distribution with \textbf{content bias}. Alternative content biases extended from unsupervised learning, supervised learning and self-supervised learning are explored in this portion of study. The VC experiments on VCTK dateset demonstrate a clear stabilized vocalization and a significantly improved performance with the new content embeddings. Phoneme classification with $z_c$ also justifies the effectiveness of the proposed model in an objective way.  

\begin{figure*}[t]
    \centering
    \includegraphics[width=0.82\linewidth, trim=0cm 1.5cm 0.5cm 0.cm height=4.0cm]{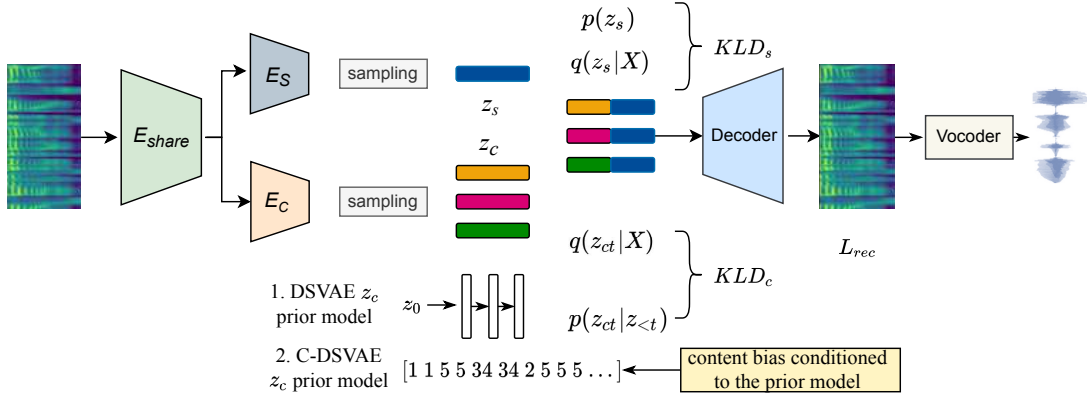}
    \vspace{-2ex}
    \caption{The system diagram of conditional DSVAE.}
    \label{dsvae fig}
\end{figure*}
   \vspace{-1ex}
\section{The DSVAE Baseline}
\subsection{Related Work}
DSVAE~\cite{dsvae} was proposed as a sequential generative  model that disentangles the time-invariant information from the time-variant information in the latent space. The original DSVAE model~\cite{dsvae} and its early variants~\cite{s3vae, c-dsvae} achieved limited success in speech disentanglement. Recently, we extended the DSVAE by balancing the information flow between speaker and content representations and it achieved the state-of-the-art performance for zero-shot non-parallel VC~\cite{D-DSVAE}. To be unified, we refer to DSVAE as the baseline we developed in~\cite{D-DSVAE}, although it is already very different from the previous systems~\cite{dsvae,s3vae,c-dsvae}. 
\subsection{Baseline Overview}
The DSVAE baseline adopted here is shown in Fig.~\ref{dsvae fig}. Denote $X, \hat{X}, z_s, z_s, \theta$  as input melspectrogram, reconstructed melspectrogram, speaker embedding, content embedding and model parameters, respectively. The shared encoder $E_{share}$ takes $X$ as input and outputs a latent representation, with the speaker encoder $E_S$ and the content encoder $E_C$ modeling the posterior distribution $q_{\theta}(z_s|X)$ and $q_{\theta}(z_c|X)$ subsequently. $z_s$ and $z_c$ are then sampled from  $q_{\theta}(z_s|X)$ and $q_{\theta}(z_c|X)$. In the next stage, the decoder takes the concatenation of $z_s$ and $z_c$, and passes them into decoder $D$ to reconstruct the melspectrogram $\hat{X}$, i.e. $\hat{X}=D(z_s, z_c)$. The vocoder then converts $\hat{X}$ into waveform. Both the prior distribution $p_{\theta}(z)$ and the posterior distribution $q_{\theta}(z|X)$ are designed to follow the independence criterion, which is similar to~\cite{dsvae, s3vae, c-dsvae, D-DSVAE}. Specifically, they can be factorized as Eq.~(\ref{prior independence}) and Eq. (\ref{posterior independence}). Note that we use $q_{\theta}(z_{ct}|X)$ to model the content posterior since the content encoder consists of BiLSTM modules, which is slightly different from the streaming posterior $q_{\theta}(x_{ct}|X{<t})$ described in \cite{dsvae, s3vae},  where they adopt unidirectional LSTM or RNN.
\vspace{-1ex}
\begin{equation} \label{prior independence} 
\resizebox{0.91\hsize}{!}{
    $p_{\theta}(z_s,z_c)=p(z_s)p_{\theta}(z_c)=p(z_s)\prod_{t=1}^T p_{\theta}(z_{ct}|z_{c}{<t})$}
\end{equation}
\vspace{-3.5ex}
\begin{equation} \label{posterior independence}
\resizebox{0.91\hsize}{!}{
         $q_{\theta}(z_s,z_s|X)=q_{\theta}(z_s|X)q_{\theta}(z_c|X)=q_{\theta}(z_s|X)\prod_{t=1}^T q_{\theta}(z_{ct}|X)$}
\end{equation}
\vspace{-4ex}
\subsection{Training and Inference}
During training, the model takes fixed length of $X$ as input and optimizes with three objectives: $\mathcal{L}_{REC}$, $\mathcal{L}_{KLD_s}$ and $\mathcal{L}_{KLD_c}$, as Eq.~(\ref{recloss}) (\ref{kls}) (\ref{klc}). $\mathcal{L}_{REC}$ is the reconstruction loss, which is implemented as the negative log likelihood. $\mathcal{L}_{KLD_s}$ and $\mathcal{L}_{KLD_c}$ denote the KL divergence for speaker and content respectively. 
\vspace{-1ex}
\begin{equation}\label{recloss}
    \mathcal{L}_{REC}=\mathbb{E}_{p(X)}\mathbb{E}_{q_{\theta}(X|z_{s}, z_{c})}[-log(q_{\theta}(X|z_s,z_c))]
\end{equation}
\vspace{-3.5ex}
\begin{equation}\label{kls}
    \mathcal{L}_{KLD_s}=\mathbb{E}_{p(X)}[ KLD(q_{\theta}(z_s|X)||p(z_s))]
\end{equation}
\begin{equation}\label{klc}
    \mathcal{L}_{KLD_c}=\mathbb{E}_{p(X)}[KLD(q_{\theta}(z_c|X)||p_{\theta}(z_c))]
\end{equation}
Given $X_1$ as the source utterance and $X_2$ as the target utterance for VC inference, the transferred sample is simply $D(z_{s2}, z_{c1})$, where $z_{s2}$ and $z_{c1}$ are sampled from $q_{\theta}(z_s|X_2)$ and $q_{\theta}(z_c|X_1)$. We use a vocoder to convert the mel spec to the waveform. 
\subsection{Implementation Details}\label{model details}

Table.~\ref{dsvae-baseline} provides detailed descriptions of each module of the DSVAE baseline. For shared encoder and decoder, the instance normalization~\cite{instance} is applied on both time and frequency axis. For speaker encoder $E_S$, content encoder $E_C$ and the content prior model $p_{z_c}$, two dense layers are used to model the mean and standard deviation of the $q(z_s|X)$, $q(z_{ct}|X)$, $p(z_{ct}|z_c{<t})$ respectively. For the prior models, $p(z_s)$ is the standard normal distribution and $p_{\theta}(z_c)$ is modeled by an autoregressive LSTM: at each time step $t$, the model generates $p(z_{ct}|z_c{<t})$, from which $z_{ct}$ is sampled and taken as the input for next time step. Note that $p_{\theta}(z_c)$ is independent of the input data $X$. The decoder consists of a prenet and postnet, which is introduced in~\cite{autovc}. We use HiFi-GAN V1~\cite{hifigan} instead of WaveNet ~\cite{wavenet} as vocoder since HiFi-GAN results in better speech quality with much faster inference speed. The vocoder is pretrained with VCTK~\cite{vctk2017} and is not involved in the training. 
\begin{table}[!ht]
\centering
\begin{adjustbox}{width=220pt,center}
\begin{threeparttable}
\begin{tabular}{|c c|} 
 \hline
\multicolumn{2}{|c|}{\textbf{Shared Encoder $E_{share}$}} \\
\multicolumn{2}{|c|}{(Conv1D(256, 5, 2, 1)$\rightarrow$ InstanceNorm2D$\rightarrow$ ReLU)$\times$3} \\
 \hline\hline
 \textbf{Speaker Encoder $E_S$} &\textbf{Content Encoder $E_C$} \\ 
 BiLSTM(512, 2)$\rightarrow$ Pooling & BiLSTM(512, 2)$\rightarrow$RNN(512, 1) \\
Dense(64)  & Dense(64)\\ 
 \hline\hline
 \multicolumn{2}{|c|}{\textbf{Decoder-PreNet $D_{Pre}$}} \\
\multicolumn{2}{|c|}{(InstanceNorm2D$\rightarrow$ Conv1D(512, 5, 2, 1)$\rightarrow$ ReLU)$\times$3} \\
\multicolumn{2}{|c|}{LSTM(512, 1) $\rightarrow$ LSTM(1024, 2) $\rightarrow$ Dense(80)} \\
 \hline\hline
 \multicolumn{2}{|c|}{\textbf{Decoder-PostNet $D_{Post}$}} \\
\multicolumn{2}{|c|}{(Conv1D(512, 5, 2, 1)$\rightarrow$  tanh$\rightarrow$ InstanceNorm2D)$\times$4} \\
 \hline\hline
 \multicolumn{2}{|c|}{\textbf{Vocoder $D$}: HiFiGAN-V1} \\
  \hline\hline
   \textbf{Prior $p(z_c)$ } &\textbf{Prior $p(z_s)$} \\ 
    LSTM(256, 1)$\rightarrow$Dense(64) & N(0,I) \\
 \hline
\end{tabular}
\caption{Detailed DSVAE architecture. For Conv1D, the configuration is (output channels, kernel size, padding, stride). For LSTM/BiLTSM/RNN, the configuration is (hidden dim, layers). For Dense layer, the configuration is (output dim).}
\label{dsvae-baseline}
\end{threeparttable}
\end{adjustbox}
\end{table}

\vspace{-6ex}

\section{Conditional DSVAE}
\subsection{Conditional Prior Distribution}
Ideal disentanglement requires $z_s$ to carry speaking style information and $z_c$ to carry content information without losing the phonetic structure. One problem for the vanilla DSVAEs~\cite{dsvae, s3vae, c-dsvae, D-DSVAE} is that the prior distribution is randomly initialized, thus it does not impose any constraint to regularize the posterior distribution. We argue that such randomness on the content prior distribution $p_{\theta}(z_c)$ impedes the content embedding $z_c$ from learning the phonetic structure information. Since the phonetic structure is explicitly modeled by $q_{\theta}(z_c|X)$, according to Eq.~\ref{klc}, one of the objective is to minimize the KL divergence between $q_{\theta}(z_c|X)$ and $p_{\theta}(z_c)$. Thus, we expect that content embedding will be significantly influenced by the prior $p_{\theta}(z_c)$ during VAE training.  In that sense, the learned phonetic structure $q_{\theta}(z_c|X)$ for all utterances will also follow the prior distribution, which does not reflect the real phonetic structure of the utterance. Such phenomenon can be observed in Fig.~\ref{fig:tsne}(a) and Fig.~\ref{fig:tsne}(c) which gives the t-SNE~\cite{tsne} visualization of $z_c$ comparing the learned content embeddings from the pretrained DSVAE~\cite{D-DSVAE} and the raw melspectrogram of the same utterances. It is observable that DSVAE representations are not phonetically discrimative in comparison to melspectrogram and they actually follow the random distribution. The aforementioned problem is detrimental to disentanglement and will generate discontinuous speech with non-stable vocalizations. 
 
 Our solution is that, instead of modeling $p_{\theta}(z_c)$, we will model the conditional content prior distribution $p_{\theta}(z_c|Y(X))$ such that the prior distribution is meaningful in carrying the content information. We call $Y(X)$ as the \textit{content bias}. The expectation is that, by incorporating the content bias into the prior distribution $p_{\theta}(z_c)$, the posterior distribution $q_{\theta}(z_c|X)$ will retain the phonetic structure of $X$. 
 
\subsection{Proposed C-DSVAE}
Based on the aforementioned discussion, we introduce four conditional DSVAE candidates:  C-DSVAE(Align), C-DSVAE(BEST-RQ), C-DSVAE(Mel) and C-DSVAE(WavLM) based on different content bias source. 
\vspace{-1ex}
\paragraph*{C-DSVAE(Align)}
In order to let $z_c$ or $q_{\theta}(z_c|X)$ to keep the phonetic structure of the speech data $X$, the content bias $Y(X)$ is expected to carry the fine-grained phonetic information. One natural choice is to let $Y(X)$ be the forced alignment of $X$. To do so, we employ the Kaldi toolkit~\cite{povey2011kaldi} to train a monophone model with 42 phonemes to obtain the forced alignment. The training portion of the VCTK dataset is used in the HMM training (see Sec.~\ref{dataset} for dataset split). We denote this bias as $Y_{Align}$. As an example showing in Fig.~\ref{dsvae fig}, the content bias $Y_{Align}$ for the current utterance is the forced alignment labels [1 1 5 5 34 34 2 5 5 5]. In the next step, the one-hot vectors are derived based on these labels for each frame, and are concatenated with the original inputs of $p_{\theta}(z_c)$ at each time step so that the new content prior becomes $p_{\theta}(z_c|Y(X))$. Such conditioned content prior is still factorized in a streaming manner, which is described as Eq.~\ref{prior dsvae}. 
 \begin{equation} \label{prior dsvae} 
    p_{\theta}(z_c|Y(X))=\prod_{t=1}^{T}P_{\theta}(z_{ct}|z_{<t}, Y(X_t))
\end{equation}
Note that $Y_{Align}$ is derived in the supervised manner, which has to reply on the audio-transcription pairs. However, transcription is not always available in practical usage. We present a few unsupervised labeling methods as the content bias candidates. We note all these methods as \textit{Pseudo Labeling} (PL), which is also mentioned in~\cite{hsu2021hubert}. The essence of PL is to derive closed-set discrete acoustic units given continuous speech input. 
\paragraph*{C-DSVAE(BEST-RQ)} Given the continuous representations as input, VQ-VAE~\cite{vqvae} will derive the corresponding quantized vector as well as discrete indices by looking up in a closed-set codebook. We adopt BEST-RQ~\cite{randomvq} to extract pseudo labels. Specifically, the melspectrogram  is linearly projected into frame-wise vectors, and then nearest-neighbour search is performed within a codebook to derive pseudo labels. Both the projection matrix and codebook are randomly initialized and then fixed during training. We denote this bias as $Y_{BEST-RQ}$. 
\paragraph*{C-DSVAE(Mel)} BEST-RQ~\cite{randomvq} is more like an online clustering algorithm that generates the pseudo labels without seeing the entire dataset. In contrast, kmeans is an offline method that embraces more global information. We directly perform kmeans on the offline melspectrogram features on the whole training data. After that, the index of cluster center is used as the pseudo label. This is consistent with the first step of HuBERT~\cite{hsu2021hubert}. We denote this bias as $Y_{Mel}$. 
\paragraph*{C-DSVAE(WavLM)} The problem in C-DSVAE(Mel) is that melspectrogram is noisy and not linguistically discriminative. To handle this problem, we attempt to apply kmeans on the pre-trained features. Specifically, we use the pre-trained WavLM features for kmeans clustering~\cite{wavlm}. The advantage of WavLM is that the aforementioned bias from melspectrograms will be alleviated via iterative clustering and the masked prediction training process. The other point is that WavLM acts as a teacher model so that the phonetic structure knowledge can be transferred from a larger corpus, which potentially improves the robustness and generalization capacity. We denote this bias as $Y_{WavLM}$. We use the WavLM Base model, which is pretrained with 960 hours of Librispeech data~\cite{panayotov2015librispeech}. 

Kmeans++~\cite{kmeans++} is employed for implementing clustering. The number of cluster is set as 50 for all experiments. We still keep speaker prior $p(z_s)$ to be a Gaussian prior $p(z_s)$, which is actually a common assumption in speaker recognition.

\subsection{Training Objective}
The content conditioned KL divergence loss is shown in Eq.~\ref{klc-cond}. The overall loss is shown in Eq,~\ref{loss}, where $\alpha$ and $\beta$ are the factors that balances the disentanglement~\cite{D-DSVAE}.
\begin{equation}\label{klc-cond}
    \mathcal{L}_{KLD_{c-cond}}=\mathbb{E}_{p(X)}[KLD(q_{\theta}(z_c|X)||p_{\theta}(z_c|Y(X)))]
\end{equation}
\begin{equation}\label{loss}
    \mathcal{L}_{C-DSVAE}= \mathcal{L}_{REC}+\alpha \mathcal{L}_{KLD_s}+\beta \mathcal{L}_{KLD_c-cond}
\end{equation}
Following~\cite{D-DSVAE}, we use the same training configuration for all experiments: the ADAM optimizer is used with the initial learning rate of 5e-4~\cite{adam}. Learning rate is decayed every 5 epochs with a factor of 0.95. Weight decay is 1e-4,  the batch-size is 256. Both speaker embedding and frame-wise content embedding are 64-D. $\alpha=0.01$ and $\beta=10$ are kept the same as~\cite{D-DSVAE}.

\section{Experiments}
\subsection{Dataset} \label{dataset}
We use VCTK corpus for experimental study~\cite{vctk2017}. 90\% of the speakers are used for training and the remaining 10\% are used for evaluation~\cite{D-DSVAE}. Melspectrogram is used as acoustic feature with the window size/hop size of 64ms/16ms, and the feature dimension is 80. We randomly select segments of 100 frames (1.6s) from the whole utterances for training. 

\subsection{Experimental Results}

\subsubsection{Content embedding and phoneme Classification}\label{phoneme experiments}
\begin{figure}[htbp]
    \begin{minipage}[b]{0.3\linewidth}
        \centering
        \centerline{\includegraphics[height=2.4cm, trim=0.5cm 0.5cm 0.5cm 0.5cm,width=2.9cm]{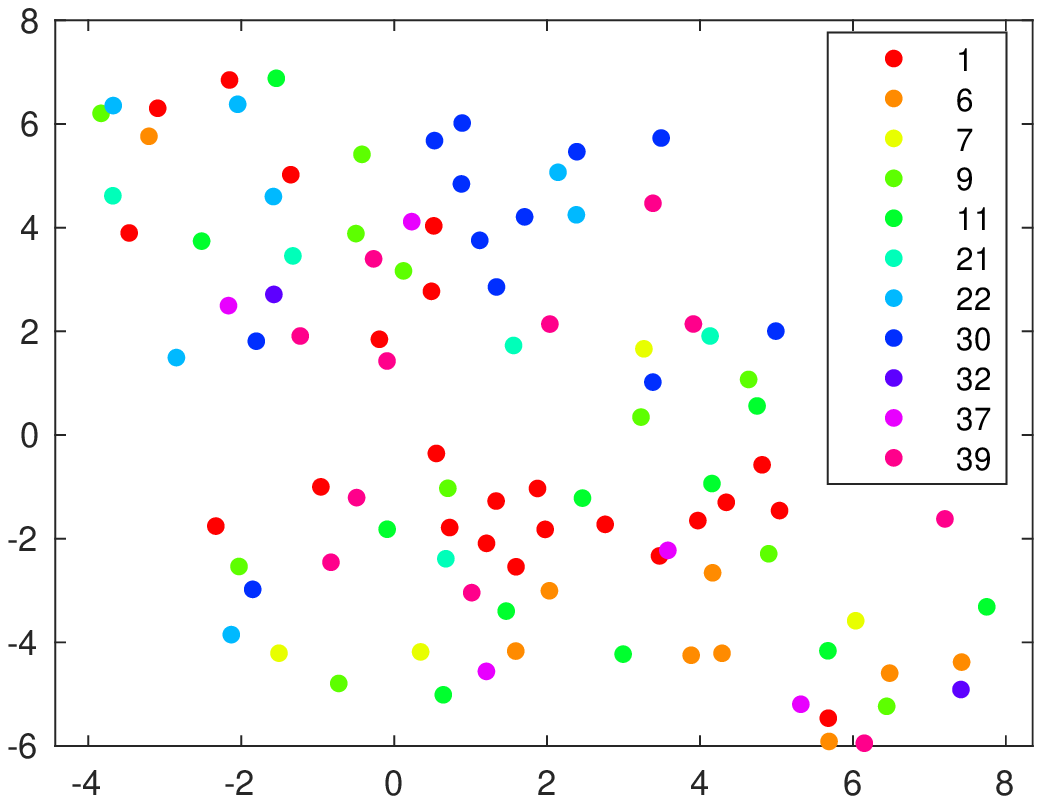}}
      %  \vspace{1.5cm}
        \centerline{\small(a) DSVAE }\medskip
      \end{minipage}
      \hfill
      \begin{minipage}[b]{0.3\linewidth}
        \centering
        \centerline{\includegraphics[height=2.4cm,trim=0.cm 0.5cm 0.5cm 0.5cm, width=2.9cm]{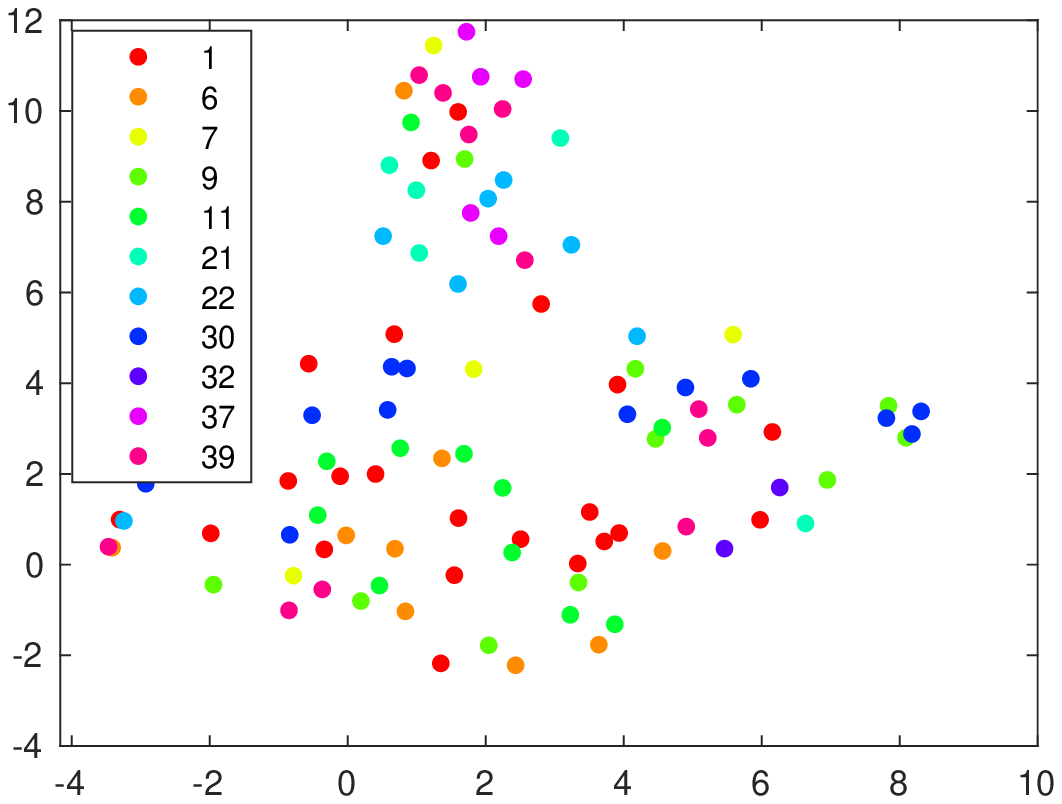}}
      %  \vspace{1.5cm}
        \centerline{(b) C-DSVAE(BEST-RQ)}\medskip
      \end{minipage}
      \hfill
      \begin{minipage}[b]{0.3\linewidth}
        \centering
        \centerline{\includegraphics[height=2.4cm, trim=0.cm 0.5cm 0.5cm 0.5cm,width=2.9cm]{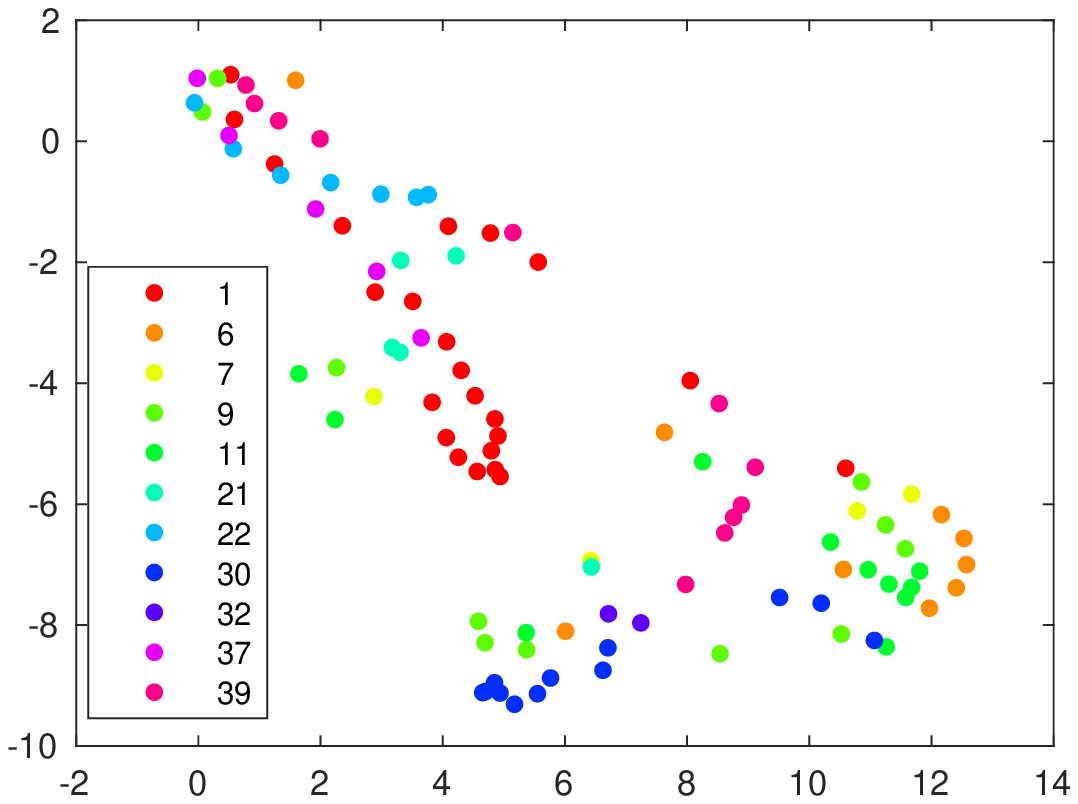}}
      %  \vspace{1.5cm}
        \centerline{\small(c) Melspec }\medskip
      \end{minipage}
            \hfill
    \begin{minipage}[b]{0.3\linewidth}
      \centering
      \centerline{\includegraphics[height=2.4cm,trim=0.5cm 0.5cm 0.5cm 0.5cm, width=2.9cm]{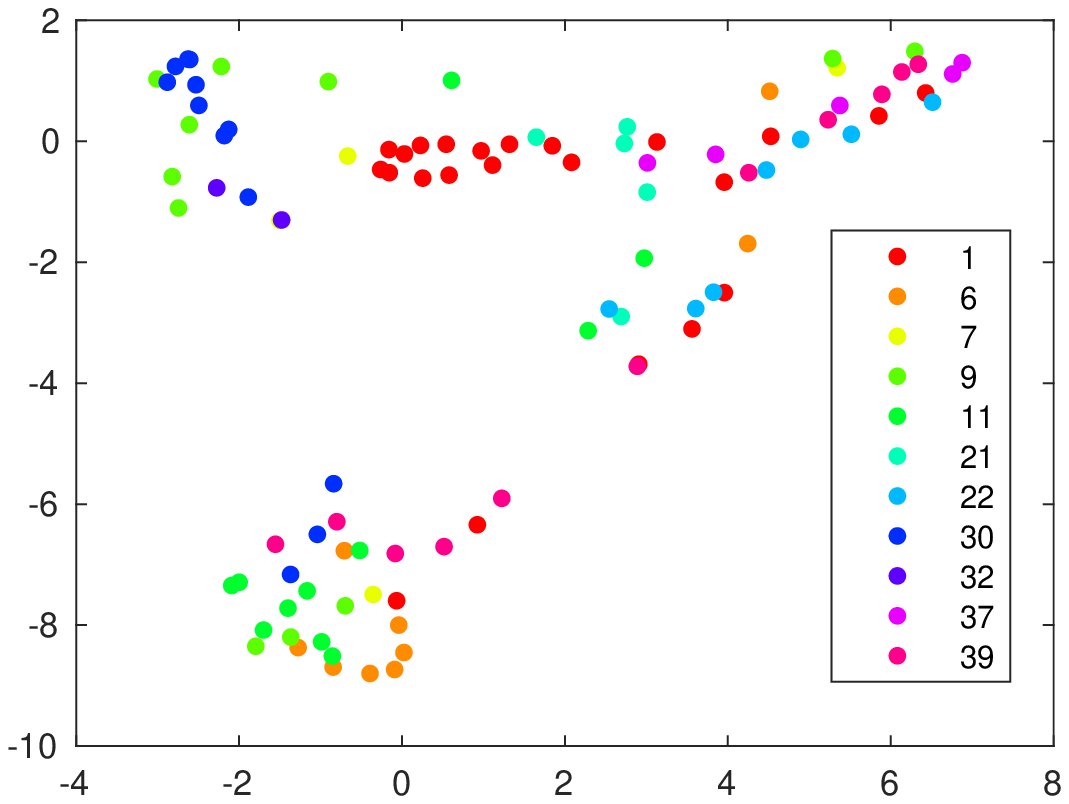}}
    %  \vspace{1.5cm}
      \centerline{(d) C-DSVAE(Mel) }\medskip
    \end{minipage}
    \hfill
    \begin{minipage}[b]{0.3\linewidth}
      \centering
      \centerline{\includegraphics[height=2.4cm,trim=0.cm 0.5cm 0.5cm 0.5cm, width=2.9cm]{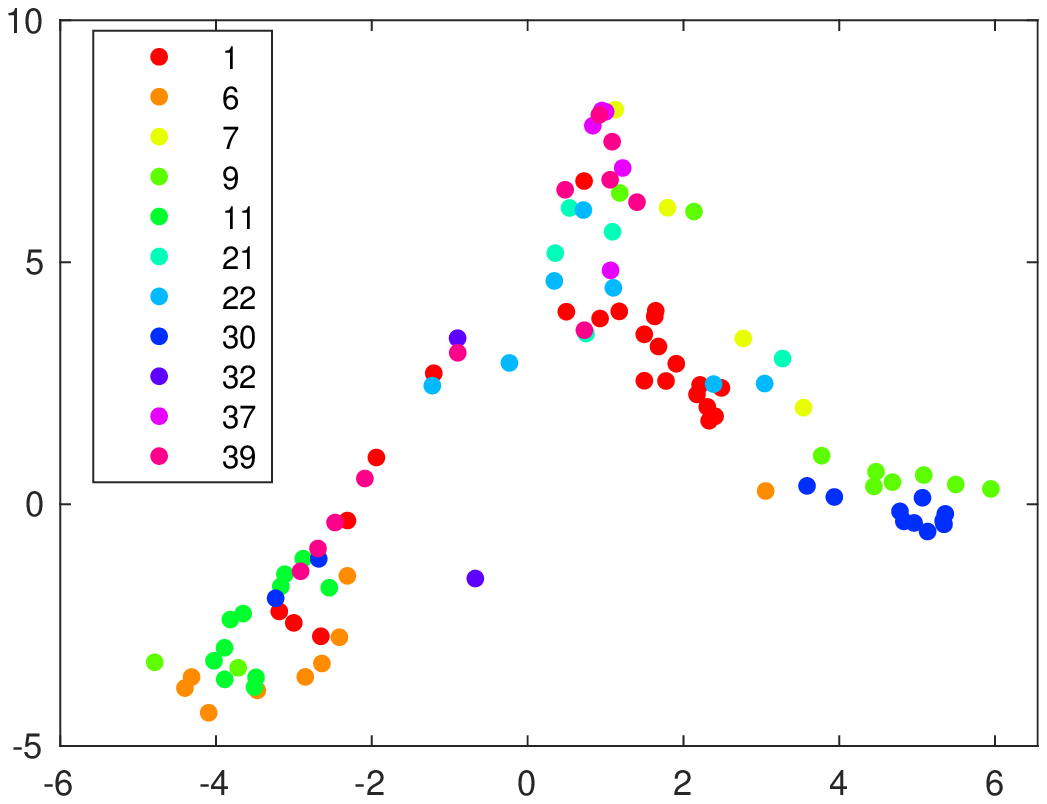}}
    %  \vspace{1.5cm}
      \centerline{(e) C-DSVAE(Align) }\medskip
    \end{minipage}
      \hfill
      \begin{minipage}[b]{0.3\linewidth}
        \centering
        \centerline{\includegraphics[height=2.4cm,trim=0.cm 0.5cm 0.5cm 0.5cm, width=2.9cm]{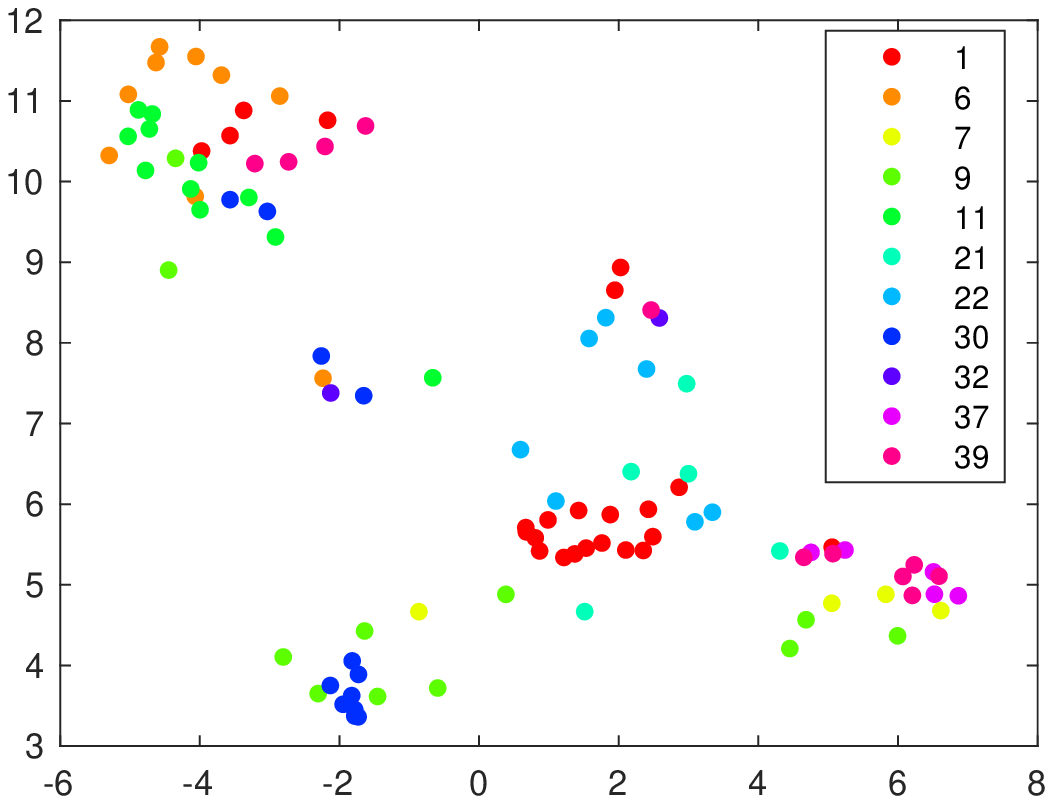}}
      %  \vspace{1.5cm}
        \centerline{(f) C-DSVAE(WavLM)}\medskip
      \end{minipage}

    \vspace{-2ex}
    \caption{Visualizations of learned content embeddings.}
    \label{fig:tsne}

\end{figure}

Fig.\ref{tab:phnclass} demonstrates the t-SNE~\cite{tsne} visualizations of the content embeddings $z_c$ from 6 different content embeddings. The purpose of this portion of study is to show how much the underlying distribution of $z_c$ matches the (almost) ground truth phonetic structure. As shown in Fig.~\ref{fig:tsne}, content embeddings from the DSVAE baseline follow a random uniform distribution. Such distribution is detrimental to preserve the phonetic structure of raw speech. At the same time, melspectrogram  captures phone-dependent information due to continuous speech signal nature. C-DSVAE(BEST-RQ) employs a random labelling process for content biasing, thus it justifies that constraint is needed for better performance. C-DSVAE(Mel), C-DSVAE(Align) and C-DSVAE(WavLM) deliver much 
desired content distributions which successfully result in phonetically discriminative embeddings. The phonetic structure of raw speech is retained and better disentanglement is expected.

We also perform phoneme classification to evaluate content embeddings in an objective way. The phoneme classifer is mentioned in Sec.~\ref{model details}. The consistent conclusion could be drawn that DSVAE and C-DSVAE(BEST-RQ) give lower accuracy. The reason for which C-DSVAE(Mel), C-DSVAE(Align) and C-DSVAE(WavLM) outperform melspectrogram is that the latter contains the coarse-grained phonetic structure which can be improved via offline clustering. C-DSVAE(Align) is better than C-DSVAE(Mel) since alignment is obtained with a supervised alignment model. C-DSVAE(WavLM) gives the best result because the masked language modeling and iterative clustering tend to capture better phonetic structure where the knowledge can also be transferred from the larger corpus. 
\vspace{-2ex}
\begin{table}[!htbp]
    \centering
     \resizebox{4.2cm}{!}{
    \begin{tabular}{|c |c|} 
     \hline
     Setting & Phn ACC \% \\ [0.5ex] 
     \hline\hline
     DSVAE & 30.2 \\ 
     C-DSVAE(BEST-RQ)& 35.6 \\
     Melspectrogram & 44.1\\
     C-DSVAE(Mel) & 48.2\\
     C-DSVAE(Align) & 51.1\\ 
     C-DSVAE(WavLM) & 52.8\\ 
     \hline
    \end{tabular}}
    \caption{Phoneme Classification with content embeddings.}
    \label{tab:phnclass}
\end{table}

\vspace{-6ex}
   \vspace{-3ex}
   
\subsubsection{Voice conversion\footnote{Samples of voice conversion can be found at https://jlian2.github.io/Improved-Voice-Conversion-with-Conditional-DSVAE.}}

We also conduct a mean opinion score (MOS) test to evaluate our system. The evaluation corpus setup is the same as~\cite{D-DSVAE}. The listener needs to give a score for each sample in a test case according to the criterion: 1 = Bad; 2 = Poor; 3 = Fair; 4 = Good; 5 = Excellent. The final score for each model is calculated by averaging the collected results. Table \ref{mos} shows the MOS results of different models. 

 As illustrated in the table, HiFi-GAN~\cite{hifigan} based DSVAE outperforms the WaveNet~\cite{wavenet} with the same acoustic features. Except for C-DSVAE(Mel), our proposed C-DSVAEs outperform the DSVAE baseline by a large margin in terms of naturalness and similarity under both seen to seen and unseen to unseen scenarios, and the MOS results are consistent with phoneme experiments as introduced in Sec~\ref{phoneme experiments}. The only exception is C-DSVAE(Mel), which achieves worse naturalness than C-DSVAE(BEST-RQ) and worse similarity than DSVAE baseline, the potential reason is that speaker embeddings learned in C-DSVAE(Mel) are not as discriminative as those in either DSVAE baseline or other C-DSVAEs.
\vspace{-1ex}
 \begin{table}[th]
     \scriptsize
         
     \centering
     %\resizebox{\columnwidth}{!}{%
     \resizebox{8cm}{!}{
     \begin{tabular}{|c||c|c||c|c|}
     \hline
   & \multicolumn{2}{c ||}{\bf{seen to seen}} & \multicolumn{2}{c |}{\bf{unseen to unseen}}  \\
    model & naturalness & similarity  & naturalness & similarity\\
     \hline 
     \hline
     AUTOVC \cite{D-DSVAE}  & 2.65$\pm$0.12  & 2.86$\pm$0.09     & 2.47$\pm$0.10  & 2.76$\pm$0.08    \\
     
     AdaIN-VC \cite{D-DSVAE} &2.98$\pm$0.09  & 3.06$\pm$0.07 & 2.72$\pm$0.11  & 2.96$\pm$0.09  \\
         
    DSVAE~\cite{D-DSVAE} & 3.40$\pm$0.07  & 3.56$\pm$0.06 & 3.22$\pm$0.09  & 3.54$\pm$0.07   \\
    \hline
    \hline
   DSVAE(HiFi-GAN) & 3.76$\pm$0.07  & 3.83$\pm$0.06 & 3.65$\pm$0.07  & 3.89$\pm$0.05   \\
    C-DSVAE(BEST-RQ) & 3.88$\pm$0.06  & 3.93$\pm$0.07 & 3.82$\pm$0.08  & 3.98$\pm$0.07   \\
    C-DSVAE(Mel) & 3.86$\pm$0.10  & 3.65$\pm$0.07 & 3.78$\pm$0.05  & 3.58$\pm$0.08   \\
    C-DSVAE(Align) & 4.03$\pm$0.04  & 4.12$\pm$0.07 & 3.93$\pm$0.06  & 4.06$\pm$0.07   \\
    C-DSVAE(WavLM) & 4.08$\pm$0.06  & 4.17$\pm$0.06 & 3.98$\pm$0.07  & 4.12$\pm$0.05   \\
     
     \hline 
     \end{tabular}
     
      }
       \caption{ The MOS (95\% CI) test on different models.}
     \label{mos} 

\end{table}

   \vspace{-5ex}
\subsubsection{Speaker verification}

We consider speaker verification as an objective measure to evaluate the VC performance. The speaker verification accuracy measures whether the transferred voice belongs to the target speaker. For this purpose, we generate 200 source-target pairs and produce 400 target trials from the test set. A state-of-the-art ECAPA-TDNN~\cite{desplanques2020ecapa} based speaker verification system is used to verify the speaker identity from the transferred voices. Please refer to~\cite{zhang2022e2e} for more details. We employ the cosine distance scoring method to perform verification and use 0.42 as the threshold, which is determined by a third-party test set~\cite{zhang2022e2e}. As shown in Table \ref{tab:SV}, voice transferred from system C-DSVAE(WavLM) achieved the best speaker verification accuracy. The trend is similar to the phoneme classification and VC MOS test, which indicates that stable content embeddings with more phonetic structure information boost the VC performance in both subjective and objective evaluations.

\begin{table}[!htbp]
    \centering
    \resizebox{4.2cm}{!}{
    \begin{tabular}{|c |c|} 
     \hline
     Setting &  ACC \% \\ [0.5ex] 
     \hline\hline
     DSVAE & 85.0 \\ 
     C-DSVAE(BEST-RQ)& 86.3 \\
     C-DSVAE(Mel) & 83.8\\
     C-DSVAE(Align) & 91.5 \\ 
     C-DSVAE(WavLM) & 92.3\\ 
     \hline
    \end{tabular}}
    \caption{Test accuracy for transferred voice verification across different models.}
    \label{tab:SV}
\end{table}
   \vspace{-4ex}
\section{Conclusion}
This paper proposes C-DSVAE, a novel voice conversion system that introduces the content bias to the prior modeling to enforce the content embeddings to retain the phonetic structure of the raw speech. The VC experiments on VCTK dateset demonstrate a clear stabilized vocalization and a significantly improved performance with the new content embeddings. With these contributions and progress, our C-DSVAE achieves state-of-the-art voice conversion performance. 
\clearpage
\bibliographystyle{IEEEtran}

\bibliography{mybib}

\end{document}